\documentclass[a4paper,english,british,conference, 10pt, final, onecolumn]{IEEEtran}
\usepackage[T1]{fontenc}
\usepackage[latin9]{luainputenc}
\usepackage{amsmath}
\usepackage{amsthm}
\usepackage{amssymb}
\usepackage{graphicx}
\usepackage{setspace}
\makeatletter

\@ifundefined{pageheight}{\let\pageheight\pdfpageheight}{}
\@ifundefined{pagewidth}{\let\pagewidth\pdfpagewidth}{}
\pageheight\paperheight
\pagewidth\paperwidth

\theoremstyle{plain}
\newtheorem{thm}{\protect\theoremname}
  \theoremstyle{definition}
  \newtheorem{defn}[thm]{\protect\definitionname}

\usepackage{color}
\usepackage{soul}

\makeatother

\usepackage{babel}
  \addto\captionsbritish{}
  \addto\captionsbritish{\renewcommand{\definitionname}{Definition}}
  \addto\captionsbritish{\renewcommand{\theoremname}{Theorem}}
  \addto\captionsenglish{}
  \addto\captionsenglish{\renewcommand{\definitionname}{Definition}}
  \addto\captionsenglish{\renewcommand{\theoremname}{Theorem}}
  
  \providecommand{\definitionname}{Definition}
\providecommand{\theoremname}{Theorem}

\begin{document}

\title{Structure Discrimination in Block-Oriented Models Using Linear Approximations:
a Theoretic Framework}

\author{J. Schoukens, R. Pintelon, Y. Rolain, M. Schoukens, K. Tiels, L. Vanbeylen, A.
Van Mulders, G. Vandersteen\\Vrije Universiteit Brussel, Department ELEC\\E-mail: johan.schoukens@vub.ac.be}

\maketitle

\emph{This is a postprint copy of: Structure discrimination in block-oriented models using linear approximations: A theoretic framework. J. Schoukens, R. Pintelon, Y. Rolain, M. Schoukens, K. Tiels, L. Vanbeylen, A. Van Mulders, G. Vandersteen.  Automatica, Vol. 53, 2015, pp. 225-234.  DOI: 10.1016/j.automatica.2014.12.045}\\

This work was supported in part by the Fund for Scientific Research
(FWO-Vlaanderen), by the Flemish Government (Methusalem), the Belgian
Government through the Inter university Poles of Attraction (IAP VII)
Program, and by the ERC advanced grant SNLSID, under contract 320378.

\begin{abstract}
In this paper we show that it is possible to retrieve structural information
about complex block-oriented nonlinear systems, starting from linear
approximations of the nonlinear system around different setpoints.
The key idea is to monitor the movements of the poles and zeros of
the linearized models and to reduce the number of candidate models
on the basis of these observations. Besides the well known open loop
single branch Wiener-, Hammerstein-, and Wiener-Hammerstein systems,
we also cover a number of more general structures like parallel (multi
branch) Wiener-Hammerstein models, and closed loop block oriented
models, including linear fractional representation (LFR) models.
\end{abstract}

\section{Introduction}

Among the many possibilities to model nonlinear dynamical systems,
block-oriented model structures became very popular (Giri and Bai,
2010; Billings and Fakhouri, 1982; Haber and Keviczky, 1999; Hunter
and Korenberg, 1986; Korenberg, 1991; Westwick and Kearney, 2003)
because these models offer a highly structured representation of the
nonlinear system, compared to other general nonlinear modeling approaches
like nonlinear state space models (Paduart \emph{et al.}, 2010) or
NARMAX models (Billings, 2013). 

\begin{figure}
\begin{centering}
\includegraphics[scale=0.5]{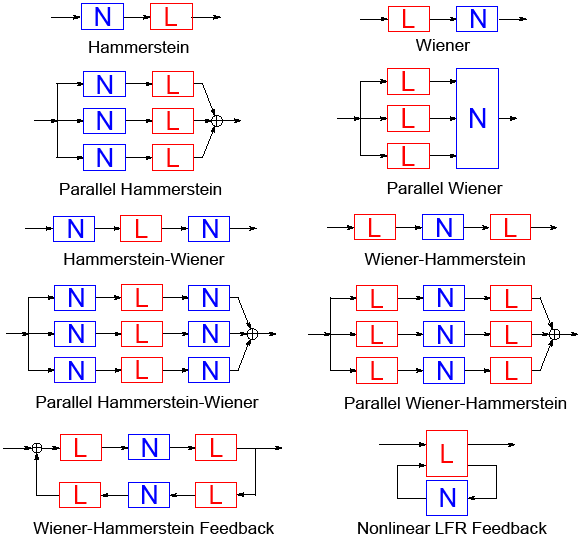}
\par\end{centering}
\caption{Some examples of block-oriented models. L stands for a linear dynamic
system, N stands for a static nonlinear system. LFR stands for a linear
fractional representation. \label{fig:All block oriented structures}}
\end{figure}

At this moment, most emphasis in the block-oriented identification
literature is on simple open loop and single branch block-oriented
models like the Wiener, Hammerstein, and the Wiener-Hammerstein or
Hammerstein-Wiener models (Giri and Bai, 2010). Although a number
of important industrial applications are reported in the literature
using these model structures, it will be shown in this paper that
their flexibility is rather limited. For that reason more general
model structures are needed to cover a wider class of nonlinear systems.
To increase the flexibility of block-oriented models, it is necessary
to consider multi path models in open and closed loop configurations
as shown in Figure \ref{fig:All block oriented structures}. This
makes it quite difficult for the user to select the best choice among
all these possibilities to tackle the problem at hand, and a lot of
time and effort can be wasted by selecting a wrong candidate model
structure at the start of the identification process.

In this paper, we look for a simple preprocessing procedure that allows
to verify if a given structure is compatible with the observations,
and this without needing to perform a full nonlinear identification.
The basic idea is to identify linear approximations of the nonlinear
system at different setpoints, and to monitor the movements of the
poles and zeros of these linear models as a function of the setpoint
changes. If a candidate model structure can not explain these movements,
it can be rejected on the basis of this information. Hence we will
provide a set of necessary but not sufficient conditions on the candidate
models. A positive test will not imply that the proposed structure
is indeed rich enough to capture the observed nonlinear behavior.
Although such a test is incomplete to do a full structure selection,
it still can save a lot of time by restricting the class of candidate
models. As such, it generalizes the initial results that were reported
by Lauwers \emph{et al.} (2008).

This paper focuses completely on the development of a theoretic formal
framework. Translating these ideas in a realistic procedure that can
be used by practicing engineers is out of the scope of this contribution. 

The paper consists mainly of two parts: in the first part (Section
\ref{sec:Class-of-excitation} and Section \ref{sec:Linearization-of-NLS})
we introduce a formal linearization framework (choice of the excitation,
choice of the linearization, study of the linearization properties
for cascaded and closed loop nonlinear systems). Next, in Section
\ref{sec:Structure-discrimination-of block oriented models}, we use
these results to obtain structural information by applying these concepts
to a number of general block-oriented model structures.

\section{\label{sec:Class-of-excitation}Class of excitation signals}

The retrieval of structural information starts from a set of linear
approximations of the nonlinear system, collected at different set
points as specified later in this paper. A linear approximation depends
strongly on the nature of the excitation signals, and the approximation
criterion that is used. In this section we discuss three classes of
excitation signals. We start with the class of Gaussian excitation
signals, extended with random phase multisines (Schoukens \emph{et
al.}, 2009). For these signals it will be illustrated in Section \ref{sec:Linearization-of-NLS}
that it is difficult to express the linearization of the complete
complex block-oriented systems (cascaded or closed loop block-oriented
systems) as a function of the linearization of the sub-systems. To
get around this problem, we will consider excitation signals that
become infinitely small, leading to the class of $\varepsilon-$bounded
excitations: signals for which the standard deviation or the maximum
amplitude is bounded by $\varepsilon$, and next we will analyze the
linearization results for $\varepsilon\rightarrow0$. The latter will
allow us to include also the classical small signal network analysis
results in the study. For these two classes of signals, it will become
possible to write the linearization of the full system as a function
of the linearization of the sub-systems. For simplicity, we will define
all the signals in the discrete time domain. It is possible to extend
the results to the continuous time domain.

\subsection{Class of Riemann equivalent Gaussian excitations}

A first class of excitation signals that we consider is the class
of Gaussian excitation signals, extended with random phase multisines.
These signals will be applied to the nonlinear system, operating around
its setpoint. To do so, a DC offset value will be added later to the
excitations, in the remainder of this section, we do not consider
this DC-offset.
\begin{defn}
Random phase multisines

Consider a power spectrum $S_{U}(f)$, that is piece-wise continuous,
with a finite number of discontinuities. A random phase multisine
is given by
\end{defn}
\begin{equation}
u(t)=\sum_{\begin{array}{c}
k=-N/2+1\\
k\neq0
\end{array}}^{N/2-1}U_{k}e^{j2\pi kt/N}\label{eq:random phase multisine}
\end{equation}
for $t=1,\ldots,N,$ and with $j^{2}=-1$. The Fourier coefficients
$U_{k}$ are either zero (the harmonic is not excited) or their amplitude
equals $\left|U_{k}\right|=\hat{U}(k/N)/\sqrt{N}$. The amplitude
function $\hat{U}(f)$ is set by the desired power spectrum $S_{U}(f)$:
\[
\hat{U}^{2}(f)=S_{U}(f).
\]
 The phases $\varphi_{k}=\angle U_{k}=-\angle U_{-k}$ are i.i.d.
such that $\mathbb{E}\{e^{j\varphi_{k}}\}=0$.

\begin{flushright}
$\square$
\par\end{flushright}

Remark: The most popular choice for the random phases $\varphi_{k}$
is to select them uniformly distributed on $[0,2\pi[$, resulting
in $\mathbb{E}\{e^{j\varphi_{k}}\}=0$. However, also a discrete distribution
can be used, for example a binary distribution $\varphi_{k}\in\{0,\pi\}$. 

Random phase multisines are asymptotically normally distributed ($N\rightarrow\infty$),
and belong to a more general class of Riemann equivalent Gaussian
excitation signals (Schoukens \emph{et al.}, 2009). Evidently, these
include also the non-periodic random excitations.
\begin{defn}
Class of Riemann equivalent excitation signals $E_{S_{U}}$\label{Defintion Class-of-Riemann equivalent signals}
\end{defn}
Consider a power spectrum $S_{U}(\omega)$, that is piece-wise continuous,
with a finite number of discontinuities. 

A random signal belongs to the equivalence class if: 

i) It is a zero mean Gaussian noise excitation with power spectrum
$S_{U}(\omega)$ 

or

ii) It is a zero mean random phase multisine s.t. 
\[
\begin{array}{cc}
\sum_{k=k_{\omega_{1}}}^{k_{\omega_{2}}}\mathbb{E}\{|U_{k}|^{2}\}=\frac{1}{2\pi}\intop_{\omega_{1}}^{\omega_{2}}S_{U}(\omega)d\omega+O(N^{-1}), & \forall k_{\omega}\end{array}
\]

with $k_{\omega_{i}}=\textrm{int(}\frac{\omega_{i}}{2\pi f_{s}}N)$,
and $0<\omega_{1},\omega_{2}<\pi f_{s}$.

\begin{flushright}
$\square$
\par\end{flushright}

It is known (Pintelon and Schoukens, 2012; Schoukens \emph{et al.},
2009) that exciting a system by any signal within a fixed class of
Riemann equivalent signals $E_{S_{U}}$, leads asymptotically ($N\rightarrow\infty)$
to the same best linear approximation.

\subsection{Class of $\varepsilon-$excitations}

Despite the fact that the class of excitation signals $E_{S_{U}}$
is very useful to deal with nonlinear systems in real life measurements,
it has some drawbacks in structure detection. As mentioned before,
we will illustrate in the next section that they are not easy to deal
with in the context of the theoretical study that is conducted here.
It will turn out that for the structure detection purpose, it is more
convenient to consider signals that become infinitely small. For such
excitation signals, the linearization of a cascade equals the cascade
of the linearization which will be the key to address the structure
analysis problem. It is clear that such a signal can not be used in
practice, but it will allow us to formalize the methods that are proposed
in this paper. Turning these results into a guidelines for the practicing
engineer remains out of the scope of this paper. 

In this section we focus on the signal $u_{\varepsilon}$, combined
with a DC-offset $u_{DC}$:

\[
u=u_{\varepsilon}+u_{DC},
\]
The excitation $u_{\varepsilon}$ operates around a setpoint $u_{DC}$,
and becomes very (infinitesimally) small as specified below. This
allows us to isolate the linear term of a system operating around
a fixed setpoint $u_{DC}$. The precise description of $u_{\varepsilon}$
is given in the next definition.
\begin{defn}
Class of $\varepsilon-$excitations $S_{\varepsilon}$

The signal $u_{\varepsilon}(t)$, $t=1,\ldots,N$, belongs to the
class of $\varepsilon-$excitations $S_{\varepsilon}$, if it belongs
to the class of Riemann equivalent excitation signals $E_{S_{U}}$
(Definition \ref{Defintion Class-of-Riemann equivalent signals}),
and 

\[
\sigma_{u}^{2}=\mathbb{E}\{u_{\varepsilon}^{2}\}=\varepsilon^{2},
\]
\end{defn}
\begin{flushright}
$\square$
\par\end{flushright}

$\varepsilon-$excitations are (asymptotically) normally distributed
signals with a limited variance. In this paper results are given for
$\varepsilon$ converging towards zero.

\subsection{Small signal analysis}

The idea of linearizing a nonlinear system around a given setpoint
is intensively used during the design and analysis of complex electronic
circuits. Popular general purpose simulation packages like SPICE (Nagels
and Pederson, 1973) offer small signal analysis options based upon
a linearization of the original system equations as used above. A
theoretic foundation for this approach can be found in the paper of
Desoer and Wong (1968), where it is shown that the solution of the
linearized nonlinear system equations comes arbitrarily close to the
small signal solution of the original nonlinear equations for excitations
with an amplitude that tends to zero. The major condition to proof
this result is that the second derivative of the nonlinear functions
exist. The approach that is developed in this paper can directly be
applied on the results of a small signal analysis. To include these
in the formal analysis that we present here, we need to add another
class of small excitation signals.
\begin{defn}
Class of $\delta-$excitations $S_{\delta}$

A signal $u_{\delta}(t)$, $t=1,\ldots,N$, belongs to the class of
$\delta-$excitations $S_{\delta}$, if 

\[
\textrm{max}|u_{\delta}(t)|=\varepsilon.
\]
\end{defn}
\begin{flushright}
$\square$
\par\end{flushright}

Remarks: 

i) We did not specify the power spectrum of $u_{\delta}$ in the previous
definition. The small signal analysis can be made using either a stepped
sine excitation, or using one of the other popular broadband excitations.
While this choice can have a strong impact on the practical aspects
of the simulation (for example the computation time), it will not
affect the theoretical results.

ii) Also the amplitude distribution is not specified, $u_{\delta}(t)$
is not requested to be Gaussian distributed as it was the case for
the class for $\varepsilon-$excitations.

iii) It might seem more logic to select $\delta$ as the amplitude
bound in the definition. However, in order to simplify the presentation
of the theorems later in this paper, we prefer to use the same upper
bound $\varepsilon$ in the definitions of $E_{\varepsilon}$ and
$E_{\delta}$ in order to simplify the formulations of the theorems
that follow later in this paper.

\section{Linearization of nonlinear systems\label{sec:Linearization-of-NLS}}

In this section we will first formally introduce the linearizations
that we consider in this paper. Next we give a brief discussion of
the related properties.

\subsection{Definitions of the linear approximations}

We focus first on Riemann equivalent excitations, as specified in
Definition \ref{Defintion Class-of-Riemann equivalent signals}. The
best linear approximation with respect to a given class of random
excitations is given by:
\begin{defn}
Best linear approximation $g_{BLA},G_{BLA}$ around a setpoint

Consider an excitation $u\in E_{S_{U}}$. The best linear approximation
around a given setpoint $u_{DC}$ is then given in the time domain
by:
\end{defn}
\begin{equation}
g_{BLA}(t)=\begin{array}[t]{c}
\textrm{arg min}\\
g
\end{array}\mathbb{E}_{u}\{(\tilde{y}(t)-g(t)*\tilde{u}(t))^{2}\}\label{eq:gBLA time domain}
\end{equation}
with $\tilde{x}(t)=x(t)-\mathbb{E}\{x(t)\}$ and $x=u$ or $x=y$
(Enqvist and Ljung, 2005; Enqvist, 2005). In the frequency domain:
\begin{equation}
G_{BLA}(\omega)=\begin{array}[t]{c}
\textrm{arg min}\\
G
\end{array}\mathbb{E}_{u}\{|\tilde{Y}(\omega)-G(\omega)\tilde{U}(\omega)|^{2}\}\label{eq:GBLA freq dom}
\end{equation}
(Pintelon and Schoukens, 2012). 

\begin{flushright}
$\square$
\par\end{flushright}

Remarks:

i) The expected value $\mathbb{E}_{u}$ is the ensemble average over
multiple realizations of the random excitation $u$. 

ii) In these expressions, $g_{BLA}$ is the impulse response of the
best linear approximation, while $G_{BLA}$ is the frequency response
function (FRF) of the best linear approximation. The dependency of
$g_{BLA},G_{BLA}$ on the setpoint $u_{DC}$ is not explicitely reflected
in the notation in order to keep the expressions simple.

iii) The best linear approximation $G_{BLA}$ is equal to the describing
function for Gaussian noise excitations, as discussed in the book
of Gelb and Vander Velde (1968) provided that these excitations operate
around the same setpoint and have a Riemann equivalent power spectrum.

The output of a nonlinear system at frequency $\omega_{k}$ can always
be written as (Pintelon and Schoukens, 2012):
\begin{equation}
\tilde{y}(t)=g(t)*\tilde{u(t)}+y_{S}(t).\label{eq:Y=00003DYBLA+YS}
\end{equation}
The first term describes that part of the output that is coherent
with the input, the second part $Y_{S}(k)$ describes the non-coherent
part. 
\begin{defn}
Local linear model $g_{\varepsilon},G_{\varepsilon}$, and $g_{\delta},G_{\delta}$
\end{defn}
Consider the best linear approximation $G_{BLA}$ obtained for a random
excitation $u=u_{DC}+u_{\varepsilon}$, and $u_{\varepsilon}\in E_{\varepsilon}$.
Define:

\begin{equation}
\lim_{\varepsilon\rightarrow0}g_{BLA}(t)\mid_{u_{\varepsilon}\in S_{\varepsilon}}=g_{\varepsilon}(t),\label{eq: defintion g-eps}
\end{equation}
and 

\begin{equation}
\lim_{\varepsilon\rightarrow0}G_{BLA}(\omega)\mid_{u_{\varepsilon}\in S_{\varepsilon}}=G_{\varepsilon}(\omega)\label{eq: defintion of G-eps}
\end{equation}

Consider the best linear approximation $G_{BLA}$ obtained for a (random)
signal $u_{\varepsilon}\in E_{\delta}$. Then we define:

\begin{equation}
\lim_{\varepsilon\rightarrow0}g_{BLA}(t)\mid_{u_{\varepsilon}\in E_{\delta}}=g_{\delta}(t),\label{eq: defintion g-delta}
\end{equation}
and 

\begin{equation}
\lim_{\varepsilon\rightarrow0}G_{BLA}(\omega)\mid_{u_{\varepsilon}\in E_{\delta}}=G_{\delta}(\omega)\label{eq: defintion of G-delta}
\end{equation}

\begin{flushright}
$\square$
\par\end{flushright}

It is possible to extend the definition of $G_{\varepsilon}$ to deterministic
signals. This is discussed and formalized in Makila and Partington
(2003), using the Fr�jet derivative. We refer the reader to this reference
for more detailed information. One of the major differences is that
the Fr�jet derivative requires the function to be differentiable,
while this is not the case for the $\varepsilon-$linearization. 

\begin{flushright}
$\square$
\par\end{flushright}

\subsection{Discussions of the properties of the linearizations}

It would be a natural choice to use the best linear approximation
$G_{BLA}$ to retrieve structural information about the block-oriented
models because it is a very convenient tool to use in practice, and
many successful applications are reported (Pintelon and Schoukens,
2012). However, it turns out that a straight forward application of
this idea to structure determination fails for a number of reasons:

i) The best linear approximation of a cascaded nonlinear systems is
not equal to the product of the best linear approximations of the
individual subsystems (Dobrowiecki and Schoukens, 2009).

ii) The best linear approximation of a closed loop system is not equal
to the closed loop calculated from the best linear approximation.

iii) For non Gaussian excitations, the best linear approximation of
a static nonlinear system can become dynamic as is illustrated in
Enqvist (2005). It is clear that this would disturb an analysis that
is based on the linearized dynamics of the nonlinear system.

For those reasons we have to use a more restricted approach that will
be offered by the $\varepsilon-$linearization or the small signal
analysis (called here $\delta-$linearization). It the following sections,
it will be shown that these linearizations have the properties that
are needed to come to a simple analysis.

\subsubsection{Linearization of a static nonlinear system}

In this section we discuss the $\varepsilon-$and $\delta-$linearization
of a static nonlinear system $y=f(u)$. First, we formalize the assumptions
on the static nonlinear system. Next, we present the linearization
properties. We use the following notations:

- the left and right limit of $f(u)$ in $u_{DC}$ is respectively
$f(u_{DC}^{-})$ and $f(u_{DC}^{+})$

- the left and right derivative of $f(u)$ with respect to $u$ in
$u_{DC}$ is $f'(u_{DC}^{-})$ and $f'(u_{DC}^{+})$.

\emph{Assumption}
Consider a static nonlinear system $y=f(u)$. In the setpoint $u_{DC}$,
we make one of the following three sets assumptions\label{thm:Assumption static nonlinear system}:

i) $f(u)$ is discontinuous in $u_{DC}$ with $f(u_{DC}^{-})=y_{DC}$,
and $f(u_{DC}^{+})=y_{DC}+c$. The left and right derivatives with
respect to $u$ can be different and are respectively: $f'(u_{DC}^{-})$
and $f'(u_{DC}^{+})$.

ii) $f(u)$ is continuous in $u_{DC}$ with $y_{DC}=f(u_{DC})$. The
left and right derivative with respect to $u$ are different and are
respectively: $f'(u_{DC}^{-})$ and $f'(u_{DC}^{+})$.

iii) $f(u)$ is continuous and differentiable in $u_{DC}$ with $y_{DC}=f(u_{DC})$
and derivative $f'(u_{DC})$. 

\begin{flushright}
$\square$
\par\end{flushright}
\begin{thm}
$\varepsilon-$and $\delta-$linearization of static nonlinear systems\label{thm: Theorem linearization static NL systems}

Consider a static nonlinear system $f$ around the setpoint $u_{DC}$,
excited by $u_{DC}+u_{\varepsilon}$, with $u_{\varepsilon}\in S_{\varepsilon}$
or $u_{\varepsilon}\in S_{\delta}$. Write the output as $y=f(u_{\varepsilon}+u_{DC})=y_{\varepsilon}+y_{DC}$
with $y_{DC}=f(u_{DC})$. Then the following properties hold for respectively
the $\varepsilon-$and $\delta-$linearization of $f$ around $u_{DC}$:

1) If Assumption \ref{thm:Assumption static nonlinear system}-i is
valid: the $\varepsilon-$and $\delta-$linearization do not exist,
they become infinite.

2) If Assumption \ref{thm:Assumption static nonlinear system}-ii
is valid: The $\delta-$linearization does not exist. The $\epsilon-$linearization
exist and is the mean of the left and right derivative: $f_{\varepsilon}(u_{DC})=0.5(f'(u_{DC}^{-})+f'(u_{DC}^{+}))$.
The output can be written as $y_{\varepsilon}=f_{\varepsilon}(u_{DC})u_{\varepsilon}+O(\varepsilon)$
with $\lim_{r_{\varepsilon}\rightarrow0}O(\varepsilon)/\varepsilon=c_{1},$
and $c_{1}$ a finite constant. The distribution of the output $y_{\varepsilon}$
differs from the normal distribution.

3) If Assumption \ref{thm:Assumption static nonlinear system}-iii
is valid: the $\varepsilon-$and $\delta-$linearizations \eqref{eq: defintion g-eps}
and \eqref{eq: defintion g-delta} exist and are equal to each other:
$f_{\varepsilon}(u_{DC})=f_{\delta}(u_{DC})=f'(u_{DC})$. The output
can be written as: $y_{\varepsilon}=f_{\varepsilon}(u_{DC})u_{\varepsilon}+O(\varepsilon^{2})$
with

$\lim_{\varepsilon\rightarrow0}O(\varepsilon^{2})/\varepsilon^{2}=c_{2}$,
and $c_{2}$ a finite constant. The output $y_{\varepsilon}$ is (asymptotically)
normally distributed for $u_{\varepsilon}\in S_{\varepsilon}$ and
$\varepsilon\rightarrow0$, if $|f'(u_{DC})|\geq\gamma>0$.
\end{thm}
Proof: See Appendix.

\begin{flushright}
$\square$
\par\end{flushright}

Remark 1: The $\delta-$linearization does not exist in case 2 of
the Theorem. The reason for that is that in a small signal analysis
($u_{\varepsilon}\in S_{\delta}$), the distribution of the excitation
is not specified, while the result of the linear approximation depends
on it (See Appendix 1).

Remark 2: The result of this theorem can be extended to the cascade
of two static nonlinear systems $f_{1},f_{2}$: 

a) The $\varepsilon-$linearization of $f_{2}\circ f_{1}$ is given
by the product of the $\varepsilon-$linearizations of $f_{1}$ and
$f_{2}$, provided that $f_{1}$ meets Assumption \ref{thm:Assumption static nonlinear system}-iii.
This requirement is needed in order to guarantee a Gaussian input
for the second nonlinearity. The properties of the $\varepsilon-$linearization
will then be set by what assumption is valid for $f_{2}$. 

b) The $\delta-$linearization of $f_{2}\circ f_{1}$ is given by
the product of the $\delta-$linearizations of $f_{1}$ and $f_{2}$,
provided that $f_{1}$ and $f_{1}$ meet Assumption \ref{thm:Assumption static nonlinear system}-iii.
This requirement is needed in order to guarantee that both $\delta-$linearizations
exist.

Remark 3: The results of this theorem can directly be extended to
a static nonlinearity that is sandwiched between two stable systems
$G_{1},G_{2}$. In that case the linearizations are given by the convolution
of the impulse responses $g_{1}(t),g_{2}(t)$ multiplied with the
$\varepsilon-$linearization of the static nonlinearity as described
in the previous theorem.

\subsubsection{Linearization of cascaded systems}

As discussed before, the best linear approximation of the cascade
of two nonlinear systems is not equal to the product of the best linear
approximations (Dobrowiecki and Schoukens, 2009). However, the same
does not hold true for the $\varepsilon-$and $\delta-$linearization,
as shown by the following theorem. 

\begin{figure}
\begin{centering}
\includegraphics[scale=0.4]{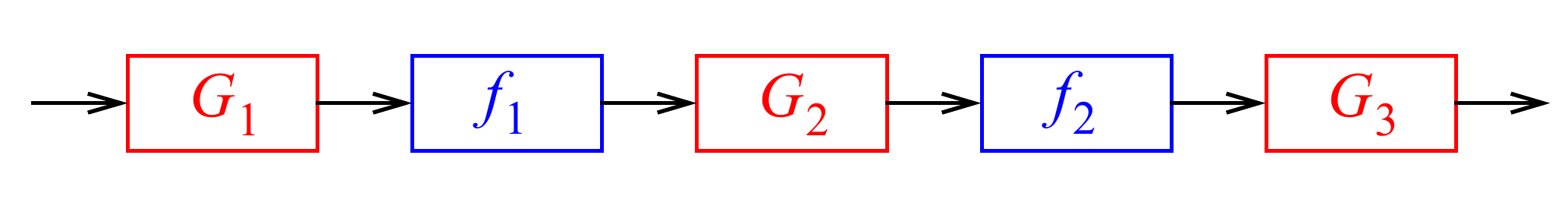}
\par\end{centering}
\caption{Cascaded single branch system\label{fig:Cascaded single branch system}}
\end{figure}

\begin{thm}
$\varepsilon-$and $\delta-$linearization of a cascaded single branch
system\label{thm: Theorem linearization-of-cascaded single branch system}

Consider the cascaded system in Figure \ref{fig:Cascaded single branch system},
excited around the setpoint $u_{DC}$ by $u_{\varepsilon}\in S_{\varepsilon}$
or $u_{\varepsilon}\in S_{\delta}$, where $\varepsilon$ indicates
the amplitude constraint of the excitation. Assume that:

- the linear systems $G_{1},G_{2},G_{3}$ are stable, 

First assume that Assumption \ref{thm:Assumption static nonlinear system}-iii
holds for $f_{1}$, while $f_{2}$ meets Assumption \ref{thm:Assumption static nonlinear system}-ii.
Then, the $\delta-$linearization of the cascaded system does not
exist. The $\epsilon-$linearization of the cascaded system $f_{\varepsilon}$
is given by the product of the $\varepsilon-$linearizations $f_{1\varepsilon}$
and $f_{2\varepsilon}$ around their respective operating points,
multiplied with the transfer function $G_{1}G_{2}G_{3}$. The difference
$e=y_{\varepsilon}-g_{\varepsilon}u_{\varepsilon}$ is an $O(\varepsilon)$. 

Next assume that both $f_{1},f_{2}$ meet Assumption \ref{thm:Assumption static nonlinear system}-iii.
Then the $\delta-$linearization and $\varepsilon-$ linearization
exist and are equal to each other. The difference $e$ is an $O(\varepsilon^{2})$,
and the output is Gaussian distributed for Gaussian excitations. 
\end{thm}
\begin{flushright}
$\square$
\par\end{flushright}

The proof follows immediately from Theorem \ref{thm: Theorem linearization static NL systems}
and the remarks following that theorem.

\subsubsection{Linearization of nonlinear feedback systems}

In this paper, we define a nonlinear feedback system as a closed loop
system with at least one static nonlinearity in the loop. We will
show in the next theorem that the $\varepsilon-$ or $\delta-$linearization
of a nonlinear feedback system is obtained by replacing the static
nonlinearity by its $\varepsilon-$ or $\delta-$linearization (Assumption
\ref{thm:Assumption static nonlinear system}-iii should be met).
Without loss of generality, we consider here a feedback loop with
a linear system $h$ in the feed-forward branch, and a single branch
feedback $q(t)=g(y(t))$ that operates around its setpoint $q_{DC}=g(y_{DC})$.
The Assumption \ref{thm:Assumption static nonlinear system}-iii is
needed for the static nonlinear blocks, to obtain a continuous and
differentiable feedback branch.
\begin{thm}
$\varepsilon-$ or $\delta-$linearization of a nonlinear feedback
system\label{thm:Feedback linearization}

Consider a (dynamic) nonlinear system $q_{\varepsilon}+q_{DC}=g(y_{\varepsilon}+y_{DC})$,
that is continuous and differentiable around its setpoint (Assumption
\ref{thm:Assumption static nonlinear system} iii). The nonlinear
system $g$ is captured in the feedback of a closed loop system:

\[
y(t)=h(t)*(r(t)-g(y(t)).
\]
The closed loop is assumed to be stable on the considered input domain
that is set by the reference signal $r=r_{\varepsilon}+r_{DC}$, with
$r_{\varepsilon}\in S_{\varepsilon}$ or $r_{\varepsilon}\in S_{\delta}$.
The value $y_{DC}$ is the response of the nonlinear system to $r=r_{DC}$.
The $\varepsilon-$ or $\delta-$linearization of the nonlinear feedback
system is then given by:

\[
Y_{\varepsilon}(\omega)=\frac{H(\omega)}{1+H(\omega)G_{\epsilon,\delta}(\omega)}R_{\varepsilon}(\omega).
\]
$H,G_{\varepsilon,\delta}$ are the transfer functions of the linear
systems $h,g_{\varepsilon}$ or $g_{\delta}$.
\end{thm}
Proof: see Appendix 2.

\begin{flushright}
$\square$
\par\end{flushright}

Remark: This result can be directly generalized to a closed loop system
with a nonlinear feed-forward and nonlinear feedback branch.

\subsubsection{Discussion}

The previous theorems show that we can replace the nonlinear sub-systems
by their $\varepsilon-$ or $\delta-$linearizations in cascaded systems
with internal nonlinear feedback loops, provided that the nonlinear
sub-systems are continuous and differentiable in their respective
setpoints (Assumption \eqref{thm:Assumption static nonlinear system}-iii).
It is important to realize that the linearization depends upon the
setpoint of the excitation, or more generally speaking, the biasing
of the system or circuit. By varying the biasing, we can vary the
linearizations of the sub-systems, resulting in a varying $\varepsilon-$
or $\delta-$linearization of the overall nonlinear system. In the
next section we will show that these variations provide information
about candidate block-oriented models that can and cannot be used
to approximate the nonlinear system.

\section{Structure discrimination of block-oriented models\label{sec:Structure-discrimination-of block oriented models}}

The main idea to retrieve structural information of a (block-oriented)
nonlinear system is to measure and model the $\varepsilon-$ or $\delta-$linearization
at a number of setpoints or bias settings, using linear system identification
tools in the time- (Ljung, 1999, S�derstr�m and Stoica, 1989) or in
the frequency domain (Pintelon and Schoukens, 2012). 

A two step procedure is proposed. First, we verify if the poles/zeros
of the linearized model vary with a changing setpoint. Next, we compare
the observed behavior with that of considered candidate model structures.
This hopefully allows many candidate block-oriented models to be excluded.

The reader should be well aware that the converse is not true: passing
the linearization test is not a sufficient condition to be sure that
the candidate structure is able to model the nonlinear systems behavior.

Three different situations are considered when evaluating the results
for different setpoints: i) all poles (zeros) remain fixed; ii) all
poles (zeros) vary; iii) some of the poles (zeros) vary with a changing
setpoint. This leads to 9 possible combinations, and for each of these
we will check if we can propose candidate model structures within
the classes of models that are considered in this paper. The discussion
will be structured along a growing complexity of the candidate model
structures.

There are a number of possibilities to change the setpoint of a system.
The first one is to apply the small excitation signals around a given
DC-level as discussed before. In that case we have to assume that
the DC-gain $G(\omega=0)$ of all the linear dynamic systems is different
from zero, so that a setpoint change affects for sure all branches
of the system. An alternative possibility for some systems is to change
the setpoints of the nonlinearities by varying the biasing of the
system using an external control signal. It is clear that also this
possibility can be used to create the varying working conditions that
are needed to make the linearization analysis. This is formalized
in the following assumption:

\emph{Assumption}
Setpoint changes\label{thm:Assumption Setpoint changes}

Consider a nonlinear block-oriented system containing the static nonlinear
functions $f_{i}$, $i=1,\ldots,n_{NL}$, operating around the setpoints
$u_{DCi}$ respectively. It is assumed that all the setpoints $u_{DCi}$
can be changed by external interactions.

In the rest of this paper, we will not deal with the dynamics of the
setpoint changes. The analysis is made under steady state conditions
for the setpoint: the setpoint is a fixed level, and after each change
of setpoint we wait till all transients in the system due to the setpoint
change become negligible.

\subsection{Single branch systems}

In a single branch system, we consider the cascade of linear dynamic
systems and static nonlinear systems. Well known members of this class
are shown in Figure \ref{fig:All block oriented structures}: the
Wiener, the Hammerstein, the Wiener-Hammerstein, and the Hammerstein-Wiener
system. It is obvious that the cascade of two single branch models
results in a new single branch model. In Wills and Ninness (2012),
more complex single branch models consisting of cascades of Hammerstein
systems are considered.

From Theorem \ref{thm: Theorem linearization-of-cascaded single branch system}
it follows that the $\varepsilon-$ or $\delta-$linearization of
the cascade is proportional to the product of the transfer functions
of all linear dynamic blocks. The gain is set by the product of the
linearizations of the static nonlinearities $f_{i}$. This leads directly
to the following theorem:
\begin{thm}
\label{thm:The-linearization-of-Single Branch system} Consider a
single branch model, consisting of a cascade of linear dynamic systems
$G_{i}$ and static nonlinear systems $f_{i}$. Assume that the systems
$f_{i}$ meet Assumptions \ref{thm:Assumption static nonlinear system}-iii
and \ref{thm:Assumption Setpoint changes}. The poles and zeros of
the $\varepsilon-$ or $\delta-$linearization do not depend on the
setpoint, and the output of the system is (asymptotically) normally
distributed for Gaussian excitations.
\end{thm}
Proof: From Theorem \ref{thm: Theorem linearization-of-cascaded single branch system},
it follows that

\begin{equation}
G_{\varepsilon}=\beta(u_{DC})\prod_{ij=1}^{n}G_{j}.\label{eq:Transfer function single branche cascade}
\end{equation}
with

\begin{equation}
\beta(u_{DC})=\prod_{j=1}^{n}\alpha_{j}(u_{DCj}),\label{eq: Overall gain single branch}
\end{equation}
where $\alpha_{i}(u_{DCi})$ is the linearization of the static nonlinear
system $f_{j}$ around its setpoint $u_{DCj}$. From \eqref{eq:Transfer function single branche cascade},
it follows immediately that the dynamics of the linearization do not
vary with the setpoint, and hence the poles and zeros do not move.

\begin{flushright}
$\square$
\par\end{flushright}

Discussion: From this theorem it follows immediately that a single
branch structure can not be used to model nonlinear systems that have
linearized dynamics that change with the setpoint: a single branch
model will not be able to capture the variations of the poles or zeros
that are observed during the linearization test. 

This result shows also the very limited capability of these models
to describe general nonlinear systems. The transfer function of the
linearized single branch model will not change dynamics for different
setpoints, only a real scaling factor will vary.

\subsection{Feed-Forward parallel systems}

In this section we consider single branch systems, put in parallel
to each other, like the parallel Wiener-Hammerstein system in Figure
\ref{fig:All block oriented structures}. In Palm (1979), it is shown
that feed-forward parallel Wiener-Hammerstein models are a very flexible
models. For that reason we include them in this study.

From Theorem \ref{thm:The-linearization-of-Single Branch system},
it follows that the precise structure of the individual branches is
not important, since the linearization for all these systems can be
written in terms of $\alpha_{i}(u_{DC})G_{i}$ if Assumption \ref{thm:Assumption static nonlinear system}-iii
is met. However, in order to make statements about the pole/zero movements
when we combine different branches, we will assume that there are
no common poles or zeros between the different branches. We also will
assume that the nonlinearities differ sufficiently from each other
in order to avoid problems with linear dependencies of the $\varepsilon-$linearizations.

The parallel structure will be linearized around a series of setpoints
$k=1,\ldots,m$ and this will result in varying gains $\beta_{i}(u_{DCk})$
\eqref{eq: Overall gain single branch} for branch $i=1,\ldots,n$
at setpoint $u_{DCk}$. Use these gains as entries for the matrix
\begin{equation}
B(k,i)=\beta_{i}(u_{DCk})\label{eq:All gain Matrix B}
\end{equation}

\emph{Assumption}
Different branches\label{thm: Assumption Different-branches}

Consider a parallel structure with $n$ parallel branches. All the
nonlinearities in a branch meet Assumption \ref{thm:Assumption static nonlinear system}-iii.

Consider the combined dynamics $G_{i}$ of the $i^{th}$ branch, being
the product of the transfer functions of the linear systems in that
branch. The systems $G_{i}$ and $G_{j}$, belonging to different
branches $i,j$, they have no common poles or zeros. 

The matrix $B$ \eqref{eq:All gain Matrix B} is of full rank.

\begin{thm}
The $\varepsilon-$ or $\delta-$linearization of a feed-forward parallel
system with n branches is given by

\begin{equation}
G_{\varepsilon}=\sum_{i=1}^{n}\beta_{i}(u_{DC})G_{i}\label{eq:Linearization //WH}
\end{equation}

Under Assumption \ref{thm: Assumption Different-branches} and \ref{thm:Assumption Setpoint changes}
for a changing setpoint $u_{DC}$, the poles of $G_{\varepsilon}$
do not move, while the zeros do.
\end{thm}
Proof: The first result \eqref{eq:Linearization //WH} follows immediately
from Theorem \ref{thm:The-linearization-of-Single Branch system}.
The second claim follows directly by replacing 
\[
G_{i}=B_{i}/A_{i}
\]
in \eqref{eq:Linearization //WH}, with $B_{i},A_{i}$ respectively
the numerator and denominator of the transfer function, giving:

\begin{equation}
G_{\varepsilon}=\sum_{i=1}^{n}\beta_{i}(u_{DC})\frac{B_{i}}{A_{i}}=\frac{\sum_{i=1}^{n}\beta_{i}(u_{DC})B_{i}\prod_{\begin{array}{c}
j=1\\
j\neq i
\end{array}}^{n}A_{i}}{\prod_{i=1}^{n}A_{i}}\label{eq://WH poles and zeros}
\end{equation}

The rational form in \eqref{eq://WH poles and zeros} has fixed poles
and moving zeros for varying setpoints.

\emph{Discussion}: 

1) Putting a number of branches in parallel increases the flexibility
of the model: the model allows to track changing dynamics by moving
the zeros as a function of the setpoint. However, it is still not
possible to move the poles. It will be necessary to create a closed
loop nonlinear model to add that flexibility as will be shown in the
next section.

2) The complexity of the feed-forward parallel model is controlled
by the prior unknown number of branches. An estimate can be obtained,
in a preprocessing step, by stacking the measured frequency response
functions $G_{\varepsilon i}$ of the $\varepsilon-$linearizations,
around the setpoints $u_{DCi},i=1,\ldots,m$ in a matrix: 
\[
G_{All}=[G_{\varepsilon1},\ldots,G_{\varepsilon m}].
\]

Under some regularity conditions, the rank of this matrix is equal
to the number of branches $n$, provided that $m\geq n$ (Schoukens
\emph{et al}., 2013; Schoukens \emph{et al}., 2015).

\subsection{Feed-forward feedback parallel systems}

Although nonlinear feedback systems can be approximated in a restricted
input domain by an open loop model (Boyd and Chua, 1985), the class
of systems allows for a much richer and more complex behavior than
the open loop system class. In this section we consider systems that
consist of a parallel feed-forward and a parallel feedback path. In
its most simple form, either the feed-forward or the feedback system
can be a linear dynamic system or a static nonlinear system. The structure
of this system will be characterized by:

- the number of branches in the feed forward ($n_{FF})$ and the feedback
($n_{FB})$, 

- the total number of branches with a static nonlinearity $n_{NL}$, 

- the total numbers of poles $n_{P}=n_{PFF}+n_{PFB}$ and zeros $n_{Z}=n_{ZFF}+n_{ZFB}$
with for example $n_{pFF}$ the number of poles in the feed-forward
and $n_{ZFB}$ the number of zeros in the feedback.

For these systems, the following theorem holds:
\begin{thm}
Under Assumption \ref{thm:Assumption static nonlinear system}-iii
and \ref{thm:Assumption Setpoint changes}, the $\varepsilon-$ or
$\delta-$linearization of a feed-forward feedback parallel system
is given by

\begin{equation}
G_{\varepsilon}=\frac{\sum_{i=1}^{n_{FF}}\gamma_{i}(u_{DC})F_{i}}{1+(\sum_{i=1}^{n_{FF}}\gamma_{i}(u_{DC})F_{i})(\sum_{j=1}^{n_{FB}}\beta_{j}(y_{DC})G_{ij})}\label{eq:Linearization //WH-1}
\end{equation}
with $F_{i},G_{j}$ respectively the linear dynamics of the $i^{th}$
feed-forward or $j^{th}$ feedback branch. Under Assumption \ref{thm: Assumption Different-branches},
and if $n_{NL}\geq1$ (at least one nonlinearity in the system), the
following results hold when the setpoint is changed:
\end{thm}
\begin{itemize}
\item All the zeros are fixed if and only if $n_{FF}=1$(single branch in
the feed-forward)
\item All the zeros move if and only if $n_{PFB}=0$ and $n_{FF}>1$ (no
poles in the feedback, more than one branch in the feed-forward)
\item All the poles are fixed if and only if $n_{FB}=0$ (no feedback)
\item All the poles move if and only if $n_{FB}\geq1$ (feedback present)
\item All the poles and the zeros are fixed if and only if $n_{FF}=1$,
and $n_{FB}=0$ (this is a single branch feed-forward system)
\item Some zeros are fixed, some zeros move if and only if $n_{FF}>1$,
$n_{PFF}+n_{ZFF}\geq1$ (a dynamic feed-forward branch), and $n_{PFB}\geq1$.
The poles of the feedback will result in fixed zeros.
\item It is not possible that some poles move and some poles are fixed
\item It is not possible that no poles move, while some zeros move and some
are fixed
\end{itemize}
Proof: The first result \eqref{eq:Linearization //WH-1} follows immediately
from Theorem \ref{thm:The-linearization-of-Single Branch system}.
The second set of claims follows directly by replacing $G_{i}=B_{i}/A_{i}$
in \eqref{eq:Linearization //WH-1}. By filling out the conditions
for each of the claims, a rational form is retrieved. The proof follows
directly by verifying if the poles and/or zeros depend on $\gamma_{i},\beta_{i}$.

\begin{flushright}
$\square$
\par\end{flushright}

\emph{Discussion}: 

1) Although a nonlinear feed-forward feedback parallel structure looks
very general, it turns out that it misses the flexibility to cover
all possible situations. In Table \ref{tab:Table //FF//FB}, an overview
is given of all possible pole/zero combinations that can be covered
by the parallel feed-forward parallel feedback structure. 

\begin{table*}
\begin{centering}
\begin{tabular}{|c|c|c|c|}
\hline 
 & $\begin{array}{c}
\textrm{all poles fixed}\\
n_{FB}=0\\
\\
\end{array}$ & $\begin{array}{c}
\textrm{some poles fixed}\\
\textrm{some poles move}\\
\textrm{NOT POSSIBLE}
\end{array}$ & $\begin{array}{c}
\textrm{all poles move}\\
\begin{array}{c}
n_{FB}\geq1\\
\\
\end{array}
\end{array}$\tabularnewline
\hline 
\hline 
$\begin{array}{c}
\textrm{all zeros fixed}\\
n_{FF}=1
\end{array}$ & single branch & $1^{*}$ & $\begin{array}{c}
\textrm{single branch FF}\\
\textrm{poles in FB}
\end{array}$\tabularnewline
\hline 
$\begin{array}{c}
\textrm{some zeros fixed}\\
\textrm{some zeros move}\\
n_{FF}\geq2,n_{FB}\geq1\\
n_{pFB}\geq1
\end{array}$ & $2^{*}$ & $3^{*}$ & $\begin{array}{c}
\textrm{multi branch FF}\\
\textrm{ poles in FB }
\end{array}$\tabularnewline
\hline 
$\begin{array}{c}
\textrm{all zeros move}\\
n_{FF}\geq2\\
n_{pFB}=0
\end{array}$ & $\begin{array}{c}
\textrm{parallel FF}\\
\textrm{no poles in FB}
\end{array}$ & $4^{*}$ & $\begin{array}{c}
\textrm{multi branch FF}\\
\textrm{ no poles in FB }
\end{array}$\tabularnewline
\hline 
\end{tabular}
\par\end{centering}
\caption{Results for nonlinear closed loop systems with a parallel feed-forward
and parallel feedback structure. The combinations $1^{*},\ldots,4^{*}$
can not be realized with this structure.\label{tab:Table //FF//FB} }
\end{table*}

From this table it is clear that some combinations can not be covered
by this structure ($1^{*},\ldots,4^{*}$). One possibility to create
such models is to cascade the feed-forward feedback parallel system
with a single branch model that will add fixed poles and zeros to
the structure. Some off these lacking combinations can also be created
with a linear fraction representation (see Section \ref{subsec: Section LFR-models})
or with other dedicated structures as will be illustrated in Section
\ref{subsec:Section special structures}.

2) It is not so easy to determine the number of branches in the feed-forward
and the feedback path. For a system with a single branch feed-forward
($n_{FF}=1)$, we have that the rank of the matrix

\[
\tilde{G}_{All}=[G_{\varepsilon1}^{.-1},\ldots,G_{\varepsilon m}^{.-1}]
\]
equals $n_{FB}+1$, with $G_{\varepsilon j}^{.-1}$ the element-wise
inverse of $G_{\varepsilon j}$. However, for a mixed structure with
$n_{FF}>1$ and $n_{FB}\geq1$, the authors are not aware of methods
that link the rank of the matrix $G_{All}$ to the number of branches
in the system.

\subsection{LFR-models\label{subsec: Section LFR-models}}

\begin{figure}
\begin{centering}
\includegraphics[scale=0.4]{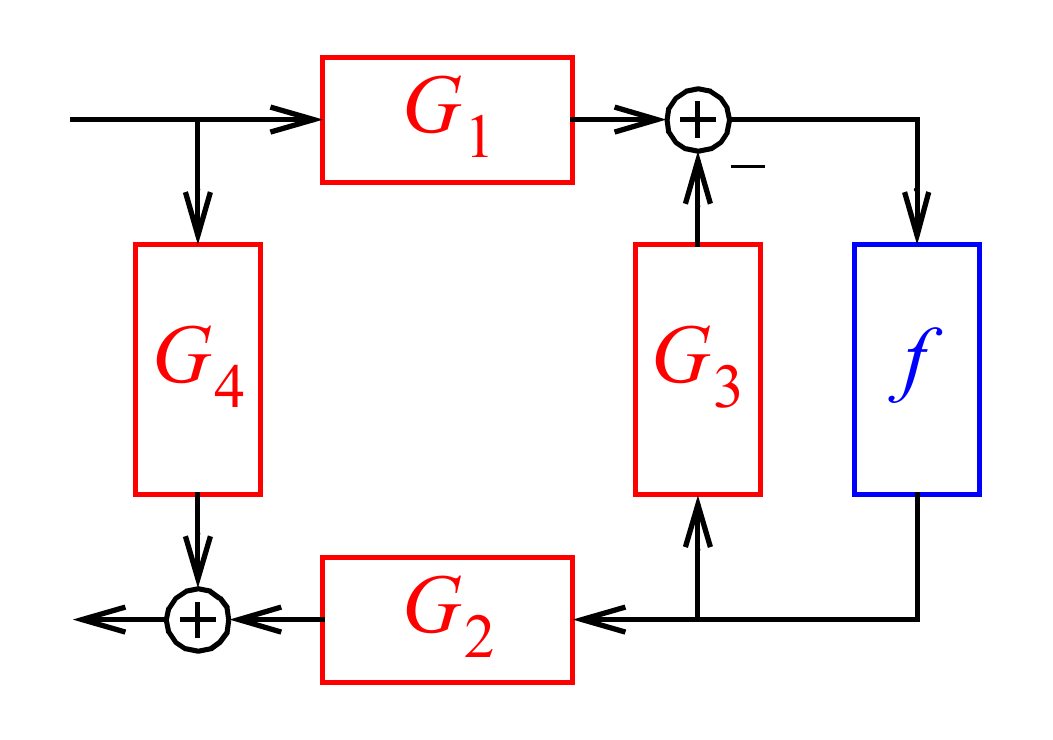}
\par\end{centering}
\caption{LFR-model\label{fig:LFR-model}}
\end{figure}

Consider the linear fraction representation (LFR) of a nonlinear system
(Vanbeylen, 2013) in Figure \ref{fig:LFR-model}.

The $\varepsilon-$linearization of this system is:
\begin{equation}
G_{\varepsilon}=G_{4}+\frac{\beta(u_{DC})G_{1}G_{2}}{1+\beta(u_{DC})G_{3}}\label{eq:LFR eps-linearization}
\end{equation}
which can be rewritten as

\begin{equation}
G_{\varepsilon}=\frac{B_{4}}{A_{4}}+\frac{\beta(u_{DC})B_{1}B_{2}A_{3}}{A_{1}A_{2}(A_{3}+\beta(u_{DC})B_{3})}\label{eq: LFR EPS-linearization poles/zeros}
\end{equation}
From \eqref{eq: LFR EPS-linearization poles/zeros}, Theorem \ref{thm: Theorem LFR structures}
follows immediately:
\begin{thm}
\label{thm: Theorem LFR structures}The $\varepsilon-$ or $\delta-$linearization
of a LFR-system that meets Assumption \ref{thm:Assumption static nonlinear system}-iii
and \ref{thm:Assumption Setpoint changes} is given by \eqref{eq: LFR EPS-linearization poles/zeros}.
Under Assumption \ref{thm: Assumption Different-branches}, for a
changing setpoint $u_{DC}$ of the input, the following results hold
for the poles and zeros:
\end{thm}
\begin{itemize}
\item If $G_{4}=0$ and provided that $G_{3}$ is not a static system: All
zeros are fixed, some poles are fixed, some poles move. The fixed
poles are those of $G_{1},G_{2}$, while the fixed zeros are those
of $G_{1},G_{2},$ and $G_{3}$. This covers situation $1^{*}$ in
Table \ref{tab:Table //FF//FB}.
\item If $G_{4}\neq0$ and provided that $G_{3}$ is not a static system:
All zeros move, some poles are fixed, some poles move. The fixed poles
are those of $G_{1},G_{2}$, and $G_{4}$. This covers situation $4^{*}$
in Table \ref{tab:Table //FF//FB}.
\end{itemize}
\emph{Discussion}: Also the LFR-structure, as defined in Figure \ref{fig:LFR-model},
misses the flexibility to cover all possible situations. It is, for
example, impossible to create a structure where some of the zeros
move, while others remain fixed.

\subsection{Checking if a structure fits with the observed movements of poles
and zeros\label{subsec:Section special structures}}

In the previous sections, we studied a number of general block-oriented
model structures. By comparing the observed pole/zero variations for
different setpoints, we can verify if a candidate model structure
is compatible with the available experimental results using linear
identification methods only, on the basis of the previous theorems.
The reader should be aware that the linearization analysis results
only in necessary conditions, it does not guarantee that the actual
structure will indeed be suitable to model that nonlinear system.
However, on the basis of the linearization results, it become possible
to propose dedicated structures that meet these necessary conditions.

\begin{figure}
\begin{centering}
\includegraphics[scale=0.4]{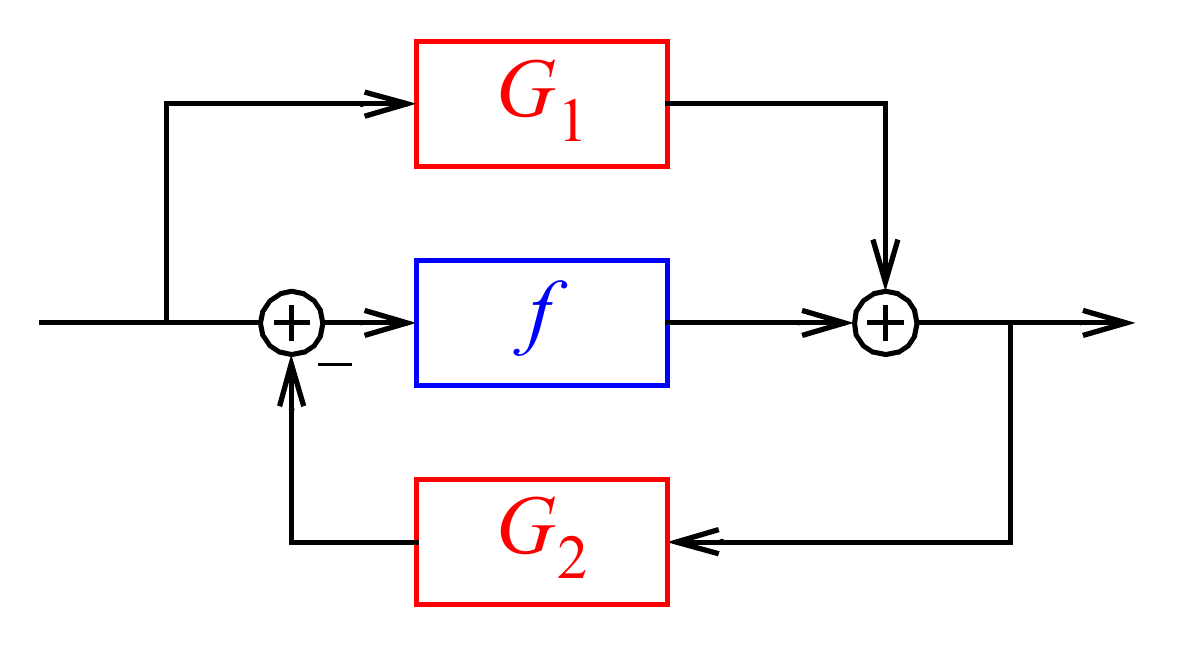}
\par\end{centering}
\caption{Symmetric FF-FB structure with a single nonlinear block. \label{fig:Symmetric FF-FB structure}}
\end{figure}

As an example, consider the structure in Figure \ref{fig:Symmetric FF-FB structure},
with
\[
G_{\varepsilon}=\frac{G_{1}+\gamma(u_{DC})}{1+\gamma(u_{DC})G_{2}}=\frac{A_{2}(B_{1}+\gamma(u_{DC})A_{1})}{A_{1}(A_{2}+\gamma(u_{DC})B_{2})}.
\]

\begin{figure}
\begin{centering}
\includegraphics[scale=0.4]{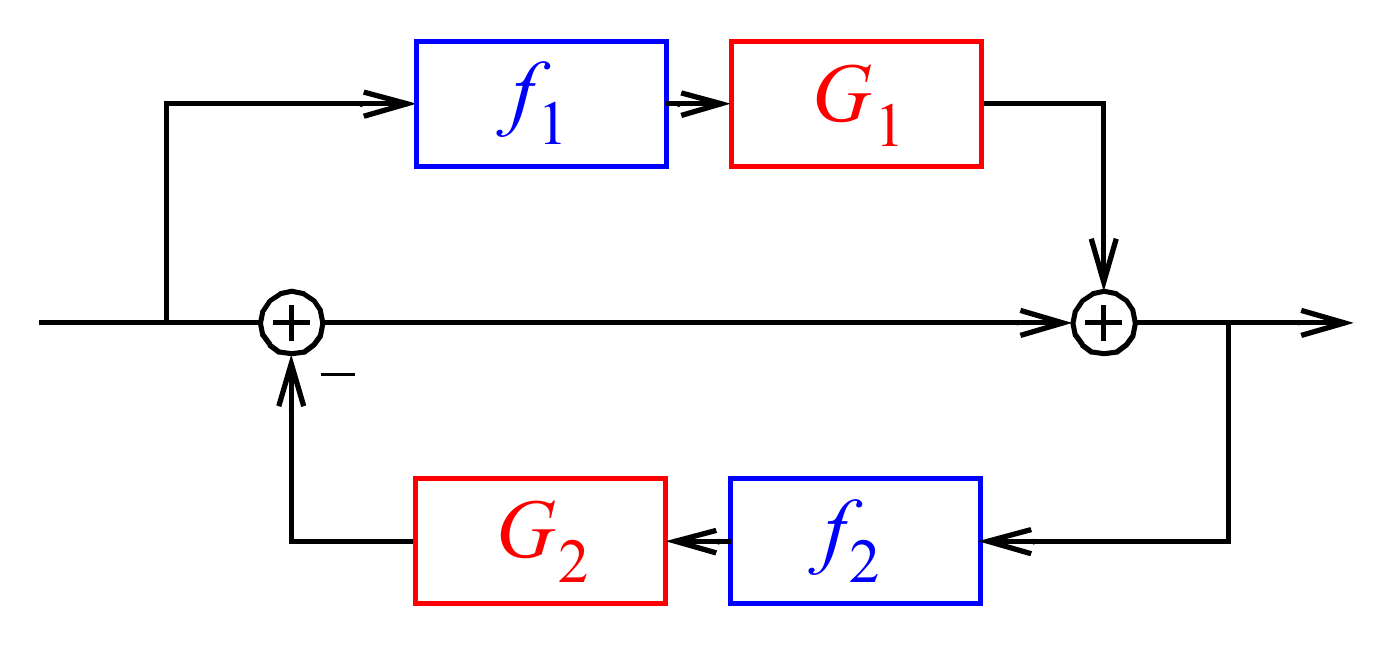}
\par\end{centering}
\caption{Symmetric FF-FB structure with two nonlinear and two linear blocks.
\label{fig:Symmetric FF-FB structure with two linear blocks}}
\end{figure}

Such a structure allows for fixed and moving poles and zeros, where
the movement is controlled by one parameter $\alpha$ that depends
on the nonlinear system $f$ and the setpoint of the system. This
covers situation $3^{*}$ in Table \ref{tab:Table //FF//FB}. More
flexibility can be created by the structure that is shown in Figure
\ref{fig:Symmetric FF-FB structure with two linear blocks}.

In that case

\[
G_{\varepsilon}=\frac{1+\gamma_{1}(u_{DC})G_{1}}{1+\gamma_{2}(u_{DC})G_{2}}=\frac{A_{2}(A_{1}+\gamma_{1}(u_{DC})B_{1})}{A_{1}(A_{2}+\gamma_{2}(u_{DC})B_{2}},
\]
which allows the movement of the poles to be decoupled from that of
the zeros by properly tuning the nonlinearities $f_{1},f_{2}$. It
is clear that both dynamic systems $G_{1},G_{2}$ can be split over
two sub-systems that can be put on the left and right side of the
corresponding nonlinearity, without changing $G_{\varepsilon}$. In
that case we have Wiener-Hammerstein systems in both nonlinear branches.
It is also possible to add a third linear system to the middle branch,
which will increase the flexibility of the pole and zero positions
even more.

\section{Simulation example}

In this simulation we give an illustration of the results obtained
in this paper. A system with two Wiener systems (S1, S2) in parallel
in the feed forward, and a Wiener system (S3) in the feedback is considered.
The system is excited with a filtered random noise excitation with
a standard deviation of 0.01 and a varying setpoints between 0 and
1 in steps of 0.1. Once the initial transients are vanished, 4096
samples are processed using the output error method (Ljung, 1999). 

The transfer function of the linear dynamic part of the three systems
is given below:

$G_{1}(z)=\frac{0.15+0.1z^{-1}}{1-0.9z^{-1}},G_{2}(z)=\frac{0.12+0.11z^{-1}}{1-0.77z^{-1}},G_{3}(z)=z^{-1}\frac{0.2+0.15z^{-1}}{1-0.72z^{-1}}.$

The delay in $G_{3}$is added in order to avoid an algebraic loop
in the system, so that it can be easily simulated using recursive
calculations.

The three static nonlinearities are respectively:

$f_{1}(x)=x-0.3x^{3},f_{2}(x)=x+0.5x^{2}+0.5x^{3},f_{3}(x)=x+0.2x^{2}+0.8x^{3}.$

\begin{figure}
\begin{centering}
\includegraphics[scale=1]{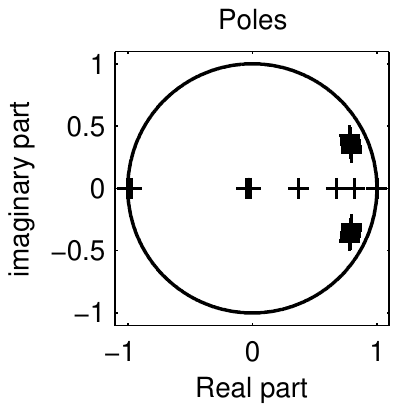}\includegraphics{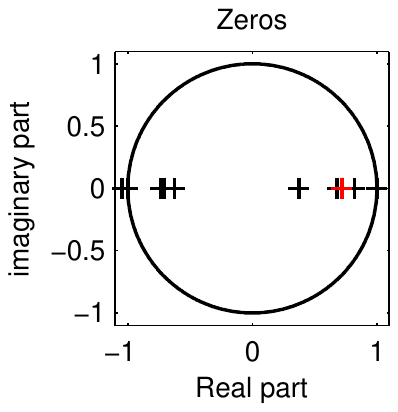}
\par\end{centering}
\caption{The poles (left) and zeros (right) of the estimated $G_{BLA}$for
the different setpoints. The estimated 'fixed' zero is given in red,
all the others are in black. \label{fig:Moving Poles and Zeros}}
\end{figure}

From Table \ref{tab:Table //FF//FB}, it follows that all poles will
move, some zeros will be fixed, and some will be move. It is easy
to verify that in this case the fixed zero will be the pole of system
$G_{3}$. The results for a simulation without disturbing noise are
given in Figure \ref{fig:Moving Poles and Zeros}. From this figure
it is clearly visible that all the poles move, while some of the zeros
are fixed. These are at the expected position.

Remarks:

i) Although there is no disturbing noise added to the output in the
simulation, the results will vary over successive realizations of
the input, because the estimated linear approximation will be influenced
by the stochastic nonlinearities. Adding noise will not change this
picture.

ii) In practice, the choice of the level of the excitation and the
varying offset levels will be a critical issue. We advice the reader
to cover the input range of interest with the different offsets that
are applied. In order to tune the excitation level, the nonparametric
nonlinear detection methods that are explained in Pintelon and Schoukens
(2012) can be used. The signal should be selected such that the signal-to-noise
ratio is as high as possible, while at the same time the observed
nonlinear distortion levels should made as small as possible in order
to meet the assumptions underlying the theory as good as possible.

iii) By making good user choices for the input signal, it is possible
to increase the quality of the best linear approximation estimates
(Pintelon and Schoukens, 2012).

iv) Till now, we considered special designed experiments around a
number of setpoints to directly identify the local linear models.
In Bai (2010), a semi-parametric local linear modeling technique is
proposed to identify nonlinear systems. Around each 'working point'
a local ARX model is identified. The local neighbourhood where each
of these ARX models is valid is set by a well designed kernel function.
The results of this paper can also be applied to the poles and zeros
of these local ARX models. This can be an interesting alternative
for the dedicated experiments that were proposed before to obtain
the linearizations. 

\section{Conclusions}

This paper proposes a set of necessary conditions on the structure
of candidate block-oriented models for a nonlinear system. To verify
these, the dependency of the poles and zeros of linear approximations
on setpoint changes is analyzed. Although it is not possible to propose
on the basis of this information a model structure that is guaranteed
to include the true system structure (no sufficient conditions), it
is possible to rule out candidate structures in an early phase of
the identification process. The results are illustrated on a number
of popular block-oriented model structures.

\subsection*{Appendix 1: Proof of Theorem \ref{thm: Theorem linearization static NL systems}}

We give the proof for $u_{\varepsilon}\in S_{\varepsilon}$, and we
will add additional remarks where needed for $u_{\varepsilon}\in S_{\delta}$.

- It is known from the Bussgang theorem, that the best linear approximation
of a static nonlinear system, excited with Gaussian noise ($u\in S_{\varepsilon}$),
is also a static system, hence the definition \eqref{eq:gBLA time domain}
reduces to:

\begin{equation}
f_{BLA}=\begin{array}[t]{c}
\textrm{arg min}\\
g
\end{array}\mathbb{E}_{u}\{(y_{\varepsilon}(t)-gu(t))^{2}\},\label{eq:gBLA static system}
\end{equation}

with $g$ a constant. The solution of \eqref{eq:gBLA static system}
is:

\begin{equation}
f_{BLA}=\mathbb{E}\{y_{\varepsilon}(t)u_{\varepsilon}(t)\}/\mathbb{E}\{u_{\varepsilon}^{2}(t)\}\label{eq: gBLA solution for static nonlinear system}
\end{equation}

- We have that:

\[
\begin{array}{c}
y_{\varepsilon}=f'(u_{DC}^{-})u_{\varepsilon}+O(\varepsilon^{2}),\textrm{if }u_{\varepsilon}\leq0\\
y_{\varepsilon}=c+f'(u_{DC}^{+})u_{\varepsilon}+O(\varepsilon^{2}),\textrm{if }u_{\varepsilon}>0
\end{array}
\]

- For $\varepsilon\rightarrow0$, the expected value $E\{yu_{\varepsilon}\}$
can be written as the sum of the contributions for $u_{\varepsilon}\leq0$
and $u_{\varepsilon}>0$.

\begin{multline*}
\mathbb{E}\{yu_{\varepsilon}\}=\\
\mathbb{E}\{f'(u_{DC}^{-})u_{\varepsilon}^{2}+O(\varepsilon^{3})\mid u_{\varepsilon}\leq0\}+\\
\mathbb{E}\{cu_{\varepsilon}+f'(u_{DC}^{+})u_{\varepsilon}^{2}+O(\varepsilon^{3})\mid u_{\varepsilon}>0\}
\end{multline*}

This expression reduces to

\begin{multline*}
\mathbb{E}\{yu_{\varepsilon}\}=\\
\mathbb{E}\{cu_{\varepsilon}\mid u_{\varepsilon}>0\}+\mathbb{E}\{f'(u_{DC}^{-})u_{\varepsilon}^{2}\mid u_{\varepsilon}\leq0\}+\\
\mathbb{E}\{f'(u_{DC}^{+})u_{\varepsilon}^{2}\mid u_{\varepsilon}>0\}+O(\varepsilon^{3})
\end{multline*}

Observe that 

\[
\mathbb{E}\{u_{\varepsilon}\mid u_{\varepsilon}>0\}=O(\varepsilon),
\]

and

\[
\mathbb{E}\{u_{\varepsilon}^{2}\mid u_{\varepsilon}\leq0\}=\varepsilon^{2}/2,
\]
\[
\mathbb{E}\{u_{\varepsilon}^{2}\mid u_{\varepsilon}>0\}=\varepsilon^{2}/2,
\]

\[
\mathbb{E}\{u_{\varepsilon}^{2}\}=\varepsilon^{2}.
\]

From these observations, the proof follows immediately for the different
situations:

1)\emph{ $f$ is discontinuous in $u_{DC}$, hence $c\neq0$}: 

\[
\mathbb{E}\{y_{\varepsilon}u_{\varepsilon}\}=O(\varepsilon)
\]

and

\[
f_{\varepsilon}=\lim_{\varepsilon\rightarrow0}f_{BLA}=\lim_{\varepsilon\rightarrow0}O(\varepsilon)/\varepsilon^{2}=\lim_{\varepsilon\rightarrow0}O(\varepsilon^{-1})=\infty.
\]

A similar argumentation can be used for the $\delta-$linearization:
the output remains finite while the input converges to zero.

2) \emph{f is continuous ($c=0)$, the derivative does not exist.} 

In that case we have that within an $O(\varepsilon^{3})$,

\[
f_{\varepsilon}=f'(u_{DC}^{-})\mathbb{E}(u_{\varepsilon}^{2}\mid u_{\varepsilon}\leq0)+f'(u_{DC}^{+})\mathbb{E}(u_{\varepsilon}^{2}\mid u_{\varepsilon}>0).
\]

From this result it turns out that the linearization depends upon
the distribution of the excitation. Since the distribution for the
class of signals $S_{\delta}$ is not specified, it follows that the
$\delta-$linearization does not exist. For excitations belonging
to $S_{\varepsilon}$we have that:

\begin{equation}
f_{\varepsilon}=f'(u_{DC}^{-})\frac{\varepsilon^{2}}{2}+f'(u_{DC}^{+})\frac{\varepsilon^{2}}{2}.\label{eq:BLA discontinuous derivative}
\end{equation}

This proves the first part of the second statement for the $\varepsilon-$linearization.

To prove the second part of the statement consider the difference
\begin{equation}
e(t)=y_{\varepsilon}(t)-g_{\varepsilon}u_{\varepsilon}(t).\label{eq:eps linearization error e}
\end{equation}
It follows immediately that

\[
\mathbb{E}\{e^{2}\}=(f'(u_{DC}^{-})-g_{\varepsilon})^{2}\frac{\varepsilon^{2}}{2}+(f'(u_{DC}^{+})-g_{\varepsilon})^{2}\frac{\varepsilon^{2}}{2}=O(\varepsilon^{2}).
\]

Since the power of the error term $e$ converges to zero with the
same rate as that of the linear term $g_{\varepsilon}u_{\varepsilon}$,
we have that the distribution of $y_{\varepsilon}$ will be the convolution
of a Gaussian distribution with that of the distribution of $e$,
and the latter is not normally distributed. This proves the last claim.

3) \emph{$f$ is continuous in $u_{DC}$, and the derivative in $u_{DC}$
exists.} 

In that case, $f'(u_{DC}^{-})=f'(u_{DC}^{+})=f'(u_{DC})$, and from
\eqref{eq:BLA discontinuous derivative}, it follows immediately that

\[
f_{\varepsilon}=f'(u_{DC}).
\]

The second part from the claim follows from the observation that the
derivative of $f$ exist in the operating point $u_{DC}$. A direct
consequence is that the non-Gaussian output contribution $e=O(\varepsilon^{2})$
\eqref{eq:eps linearization error e} converges faster to zero than
that of the linear term, and hence the distribution of $y_{\varepsilon}$
converges to that of $y_{\varepsilon}=f_{\varepsilon}u_{\varepsilon}$,
which is (asymptotically) normal.

For $u_{\varepsilon}\in S_{\delta}$ we can consider the linear term
of the Taylor expansion which is again given by $f'(u_{DC})$. Also
here the error term will be an $O(\varepsilon^{2})$.

\subsection*{Appendix 2: Proof of Theorem \ref{thm:Feedback linearization}}

We give the proof here for $y_{\varepsilon}\in S_{\varepsilon}$,
the proof for the $\delta-$linearization is completely similar. Since
the nonlinear system is continuous and differentiable (Assumption
\ref{thm:Assumption static nonlinear system}-iii), we have that for
a stable system, $y_{\varepsilon}\in S_{\varepsilon}$ converges to
zero when the excitation $r_{\varepsilon}$ goes to zero. It then
follows immediately from Theorem \ref{thm: Theorem linearization static NL systems}
that the output of the nonlinear system in the feedback loop is given
by:

\[
q_{\varepsilon}=g(y_{\varepsilon}+y_{DC})-y_{DC}=g_{\varepsilon}*y_{\varepsilon}+O(\varepsilon^{2})
\]
so that we can replace the feedback by its $\varepsilon-$linearization.
The proof follows then immediately by replacing the nonlinear system
by its $\varepsilon-$linearization $g_{\varepsilon}$. 



\begin{thebibliography}{References}
\bibitem[Bai (2010)]{Bai Er-Wei}Bai E.W. (2010). Non-parametric nonlinear
system identification: an asymptotic minimum mean squared error estimation.
\emph{IEEE Trans. on Automatic Control}, 55, pp. 1615-1625.

\bibitem[Billings and Fakhouri (1982)]{Billings-Fakhouri}Billings,
S.A. and S.Y. Fakhouri (1982). Identification of systems containing
linear dynamic and static nonlinear elements. \textit{Automatica},
18, pp. 15-26. 

\bibitem[Billings (2013)]{Billings book}Billings, S. A. (2013). \emph{Nonlinear
system identification : NARMAX methods in the time, frequency, and
spatio-temporal domains.} John Wiley \& Sons Ltd.

\bibitem[Bussgan (1952)]{Bussgang}Bussgang J.J. \emph{(1952). Cross-correlation
functions of amplitude-distorted Gaussian signals}. Technical Report,
216, MIT Laboratory of Electronics, 1952.

\bibitem[Boyd and Chua (1985)]{Boyd Fading memory paper 1985}Boyd
S. and L.O. Chua (1985). Fading Memory and the Problem of Approximating
Nonlinear Operators with Volterra Series. \emph{IEEE Trans. on Circuits
and Systems}, 32, pp. 1150-1161.

\bibitem[Desoer and Wong (1968)]{Desoer Linearizing nonlinear equations}Desoer,
C.A. and K.K. Wong (1968). Small-Signal Behavior of Nonlinear Lumped
Networks. Proc. of the IEEE, 61(1), pp. 14-22.

\bibitem[Dobrowiecki and Schoukens (2006)]{Dobrowiecki BLA cascaded systems}
Dobrowiecki T. and J. Schoukens (2006). Robustness of the related
linear dynamic system estimates in cascaded nonlinear MIMO systems.
23rd IEEE Instrumentation and Measurement Technology Conference, Sorrento,
Italy, pp. 117-122. 

\bibitem[Enqvist and Ljung (2005)]{Enqvist and Ljung}Enqvist M. and
L. Ljung (2005). Linear approximations of nonlinear FIR systems for
separable input processes. Automatica, 41(3), pp. 459-473.

\bibitem[Enqvist (2005b)]{Enqvist 2005 Thesis}Enqvist M. (2005).
\emph{Linear Models of Nonlinear systems}. PhD Thesis No. 985, Institute
of technology, Link�ping University, Sweden, 2005.

\bibitem[Gelb and Vandervelde (1968)]{Gelb and Vander Velde. Describing functions}Gelb,
A., and W. E. Vander Velde. Multiple-Input Describing Functions and
Nonlinear System Design. McGraw Hill, 1968.

\bibitem[Giri and Bai (2010)]{Giri and Bai booek}Giri, F. and E.W.
Bai (Eds.), (2010). \textit{Block-oriented Nonlinear System Identification}.
Springer.

\bibitem[Haber and Keviczky (1999)]{Haber and Keviczky} Haber, R.,
and Keviczky, L. (1999). \emph{Nonlinear System Identification: Input-Output
Modeling Approach}. Dordrecht: Kluwer Academic Publishers.

\bibitem[Hunter and Korenberg (1986)]{Hunter and Korenberg}Hunter
I.W. and M.J. Korenberg (1986). The identification of nonlinear biological
systems: Wiener and Hammerstein cascade models. \textit{Biological
Cybernetics}, 55, pp. 135-144.

\bibitem[Korenberg (1991)]{Korenberg}Korenberg M.J. (1991). Parallel
cascade identification and kernel estimation for nonlinear systems.
Annals of Biomedical Engineering, 19:429\textendash 455.

\bibitem[Lauwers et al. (2008)]{Lauwers NLstructure 2008}Lauwers
L., J. Schoukens, R. Pintelon, M. Enqvist (2008). A nonlinear block
structure identification procedure using Frequency Response Function
measurements. \emph{IEEE Trans. on Instrum. and Meas.}, 57, pp. 2257-2264,
2008. 

\bibitem[Ljung (1999)]{LjungBook} Ljung, L (1999), \emph{System Identification:}
\emph{Theory for the User (second edition)}. Prentice Hall, Upper
Saddle River, New Jersey.

\bibitem[Makila and Partington (2003)]{Makila 2003}M�kil� P.M., and
J.R. Partington (2003). On linear models for nonlinear systems. \emph{Automatica}
39, pp. 1-13.

\bibitem[Nagel and Pederson (1973)]{Original SPICE publication}Nagel,
L. W, and Pederson, D. O. (1973). SPICE (Simulation Program with Integrated
Circuit Emphasis), Memorandum No. ERL-M382, University of California,
Berkeley, Apr. 1973

\bibitem[Palm (1979)]{Palm}Palm, G. (1979). On representation and
approximation of nonlinear systems. Biological Cybern., 34, pp. 49-52,
1979.

\selectlanguage{british}%
\bibitem[Pintelon and Schoukens (2012)]{PintelonBoek}\foreignlanguage{english}{
Pintelon, R. and J. Schoukens (2012). \emph{System Identification.
A Frequency Domain Approach.} 2nd edition. Wiley - IEEE-press, Piscataway.}

\selectlanguage{english}%
\bibitem[Schetzen (2006)]{Schetzen 2006}Schetzen M. (2006). \emph{The
Volterra and Wiener Theories of Nonlinear Systems}. Wiley and Sons,
New York.

\bibitem[Schoukens et. al. (2008)]{Schoukens crystal detector}Schoukens
J., L. Gomme, W. Van Moer, et al. (2008). Identification of a block-structured
nonlinear feedback system, applied to a microwave crystal detector.
\textit{IEEE Trans. Instrum. and Measurem.}, 57, pp. 1734-1740. 

\bibitem[Schoukens et al. (2009)]{Schoukens RiemanEquivalence}Schoukens
J., J. Lataire, R. Pintelon, and G. Vandersteen (2009). Robustness
issues of the equivalent linear representation of a nonlinear system.
\emph{IEEE Trans. Instrum. Meas.}, vol. 58, pp. 1737-1745.

\bibitem[Schoukens et al. (2013)]{Maarten parallel}Schoukens M.,
G. Vandersteen, Y. Rolain (2013). An identification algorithm for
parallel Wiener-Hammerstein systems.\emph{ IEEE Decision and Control
Conference}, CDC 2013, pp. 4907-4912.

\bibitem[Schoukens et al. (2013)]{Maarten parallel-Automatica}Schoukens
M., A. Marconato, R. Pintelon, G. Vandersteen, Y. Rolain (2015). Parametric
identification of parallel Wiener-Hammerstein systems.\emph{ Automatica,
}51, pp. 111-122\emph{.}

\bibitem[Soderstrom and Stoica (1989)]{SoderstromBook} S�derstr�m,
T. and P. Stoica (1989). \emph{System Identification. }Prentice-Hall,
Englewood Cliffs 1989.

\bibitem[Vanbeylen (2013)]{Vanbeylen LFR 2013}Vanbeylen L. (2013).
Nonlinear LFR block-oriented model: Potential benefits and improved,
user-friendly identification method. \emph{IEEE Trans. on Instrum.
and Meas.}, 62, pp. 3374\textendash 3383, 2013. 

\bibitem[Westwick and Kearney (2003)]{Westwick and Kearney} Westwick,
D.T. and R.E. Kearney (2003). \textit{Identification of Nonlinear
Physiological Systems}. IEEE-Wiley.

\bibitem[Wills and Ninness (2012)]{Wills and Ninness (2012)}Wills
A. and B. Ninness (2012). Generalised Hammerstein-Wiener system estimation
and a benchmark application. \textit{Control Engineering Practice},
20, pp. 1097-1108. 
\end{thebibliography}
\end{document}